\documentclass[preprint,prb]{revtex4}

\usepackage{graphicx}
\usepackage{dcolumn}
\usepackage{bm}
\usepackage{amsmath}
\usepackage{multirow}

\newcolumntype{f}[1]{D{.}{.}{#1}}

\begin{document}

\title{On the shift in membrane miscibility transition temperature upon
  addition of short-chain alcohols}
\date{\today}
\author{M. Schick}
\affiliation{Department of Physics, University of Washington, Seattle,
  WA 98195}

\begin{abstract}
I consider the effect of a small concentration of a molecule, such as a 
short-chain alcohol, on the miscibility transition temperature of a giant 
plasma membrane vesicle. For concentrations sufficiently small such that 
the system can be treated as a dilute solution, the change in transition 
temperature is known to depend upon the extent of the molecule's partition 
into the coexisting liquid-disordered and liquid-ordered phases. 
Preferential partitioning into the former decreases the miscibility 
temperature, while preferential partitioning into the latter causes an 
increase. The analysis, combined with calculated values of the partition 
coefficient of saturated chains, illuminates the results of recent 
experiments on the change in miscibility transition temperatures with 
changing alcohol chain length, and makes several testable predictions.
\end{abstract}

\date{\today}

\maketitle

\section{Introduction}
It is well-known that model membranes, consisting of a ternary mixture of 
saturated lipids, unsaturated lipids, and cholesterol, exhibit a 
liquid-liquid miscibility phase transition \cite{veatch05}. The two liquid 
phases are distinguished by their composition and the extent of the 
variation, in the lipids' acyl chains, of the angle between adjacent 
carbon-carbon bonds and the membrane normal. The phase with the greater 
variation is denoted\cite{ipsen87} liquid-disordered (ld). It is rich in 
unsaturated lipid.  The phase with the lesser disorder is denoted\cite{ipsen87}
liquid-ordered (lo), and is rich in saturated lipids. Both 
phases are characterized by a diffusion constant typical of a 
two-dimensional fluid, as opposed to the much smaller one of the more 
ordered, and more dense, gel phase. Complex, cell-derived. giant plasma 
membrane vesicles also exhibit such a liquid-liquid 
transition\cite{veatch08}.
 
It has recently been shown that the introduction of short-chain alcohols 
into cell-derived giant plasma membrane vesicles affects the temperature 
of transition from a single, macroscopically uniform phase, to coexisting 
lo and ld phases \cite{gray13}.  The miscibility transition temperature 
decreases on the introduction of ethanol. If the length of the chain in 
the n-alcohols is made larger, the magnitude of the change in temperature 
increases through propanol, octanol, and decanol. With further increase in 
n however, that trend reverses, and the magnitude decreases such that 
tetradecanol, (n=14), exhibits no effect on the transition temperature. 
This behavior is interesting in light of the result that the introduction 
of cholesterol into a giant unilamellar vesicle consisting of a mixture of 
two miscible lipids causes them to undergo phase separation, that is, it 
{\em increases} the miscibility transition temperature \cite{veatch06}. 
The results of Gray et al.\cite{gray13} are not without precedent, 
however, as it was observed long ago that alcohols with $n$ less than 12 
depress the gel-liquid transition temperature \cite{pringle79}. The 
observed behavior was interpreted in terms of a thermodynamic result for a 
dilute solution, a result, derived below, that relates the temperature 
shift to the partitioning of the alcohol between the liquid and gel phase.

It is the purpose of this paper to show that if the alcohol forms a dilute 
solution in the membrane, then the change in the temperature of a 
first-order miscibility transition exhibits the same behavior with alcohol 
chain length as that observed by Gray et al. \cite{gray13}. To show this, 
I utilize a simple thermodynamic argument \cite{landau58} and the results 
of a recent calculation of the partition coefficients of single chains in 
coexisting lo and ld phases \cite{uline10}. This combination makes several 
testable predictions about the temperature change that would be observed 
were longer-chain alcohols to be employed, or larger concentrations of 
shorter-chain alcohols to be introduced. I also emphasize that the change 
in the temperature of a miscibility transition upon the introduction of an 
alcohol into a membrane containing $p$ components is not a well-defined 
quantity unless the behavior of the other $p+1$ independent thermodynamic 
variables is specified.

\section{Thermodynamics}
I first review the argument of Landau and Lifshitz\cite{landau58} 
concerning the change in the temperature of a first-order transition upon 
the introduction of a solute into a one-component membrane acting as a 
solvent. In the absence of solute, the internal energy of a bilayer with 
entropy $S$, number of solvent particles $N,$ and area $A,$ is given by
\begin{equation}
U=TS+\sigma A+\mu N,
\end{equation}
with differential 
\begin{equation}
dU=TdS+\sigma dA+\mu dN,
\end{equation}
where $T$, $\sigma$, and $\mu$ are the temperature, surface tension, and 
chemical potential respectively. Differentiating the first equation above 
and comparing with the second, one obtains the Gibbs-Duhem relation
\begin{equation}
\label{g-d}
SdT+Ad\sigma+Nd\mu=0.
\end{equation}
A convenient thermodynamic potential for the system is the Gibbs free energy
\begin{eqnarray}
\label{gibbs}
\Phi_0(T,\sigma,N)&=&U-TS-\sigma A,\nonumber\\
&=&N\mu_0(T,\sigma).
\end{eqnarray}
The potential can be calculated from the partition function
\begin{equation}
Q_0(T,\sigma,N)=\exp[-\Phi_0(T,\sigma,N)/k_BT]=\frac{1}{N!}Tr\exp[-(H(N,A)-\sigma 
A]/k_BT],
\end{equation}
where $H$ is the Hamiltonian of the system.
Now let $n_s$ molecules of solute be added to the system and consider the 
effect on the thermodynamic potential, which becomes 
$\Phi(T,\sigma,N,n_s).$ Because the $n_s$ solute particles are 
indistinguishable, the partition function becomes
\begin{equation}
Q(T,\sigma,N,n_s)=\exp[-\Phi(T,\sigma,N,n_s)/k_BT]=\frac{1}{N!n_s!}Tr\exp[-(H(N,n_s,A)-\sigma 
A]/k_BT],
\end{equation}
so that
\begin{equation}
\label{c1}
\Phi(T,\sigma,N,n_s)=\Phi_0(T,\sigma,N)+n_sk_BT\ln(n_s/e)-k_BT\ln\left[\frac{n_s!Q(T,\sigma,N,n_s)}{Q_0(T,\sigma,N)}\right],
\end{equation}
where Stirling's approximation has been used. Further the thermodynamic 
potential must be a homogeneous function of $N$ and $n_s$ of order one, 
i.e.
\begin{equation}
\label{c2}
\Phi(T,\sigma,\lambda N,\lambda n_s)=\lambda\Phi(T,\sigma,N,n_s),
\end{equation}
for arbitrary $\lambda$.
From Eqs. (\ref{gibbs}), (\ref{c1}), and (\ref{c2}) it can seen that for a 
weak, or dilute, solution, one for which $n_s<<N$, the thermodynamic 
potential must have the form, to first order in $n_s,$
\begin{equation}
\label{c3}
\Phi(T,\sigma,N,n_s)=N\mu_0(T,\sigma) +n_sk_BT\ln(n_s/eN)+n_s\psi(\sigma,T),
\end{equation}
where the function $\psi$ depends  only of $\sigma$ and $T$.
The first term is the potential in the absence of solute. The form of the 
second term, the entropy of mixing, follows from the fact that Eq. 
(\ref{c2}) requires that the logarithm depend on the ratio $n_s/N$; the 
third term from the fact that with the extensivity appearing directly in 
$n_s$, any function that it multiplies can depend only on powers of 
$n_s/N$, which would contribute to $\Phi$ terms of higher order in $n_s$, 
and on $\sigma$ and $T$.

From Eq (\ref{c3}) it follows that the chemical 
potential of the solvent is, to first order in the solvent mol fraction, 
or concentration $c\equiv n_s/N,$ given by
\begin{equation}
\label{mu-change}
\mu(T,\sigma,c)=\frac{\partial\Phi(T,\sigma,N,n_s)}{\partial N}=\mu_0(T,\sigma)-k_BTc.
\end{equation}
Note that this change in the solvent chemical potential arises solely 
from 
the entropy of the solute. Contributions to the solvent chemical potential 
from interactions between solute molecules and other molecules, solvent or 
solute, are of higher order in the solute concentration.

Consider a first-order transition from one uniform phase to two coexisting 
phases, denoted $I$ and $II.$ In the case of a pure one-component solvent, 
the condition for coexistence is that, in addition to the temperature and 
surface tension of each phase being equal, the thermodynamic potentials, 
or equivalently the chemical potentials, of each phase must also be equal
\begin{equation}
\mu_0^I(T_{0,co},\sigma_0)=\mu_0^{II}(T_{0,co},\sigma_0).
\end{equation}
This condition determines the coexistence curve 
$T_{0,co}=T_{0co}(\sigma_0).$ The transition temperature is completely 
determined by the surface tension of the two coexisting phases.

With the addition of a solute forming a dilute solution, the condition of 
the equality of solvent chemical potentials becomes, from Eq. 
(\ref{mu-change})
\begin{equation}
\mu_0^I(T,\sigma)-c_Ik_BT=\mu_0^{II}(T,\sigma)-c_{II}k_BT.
\end{equation}

The change in transition temperature on the addition of solute is obtained 
by expanding $\mu_0(T,\sigma)$ about $\mu_0(T_{co,0},\sigma_0)$. Denoting 
$T=T_{co,0}+\Delta T$ and $\sigma=\sigma_0+\Delta \sigma$ and utilizing 
Eq. (\ref{g-d}) from which $\partial\mu_0/\partial T=S/N\equiv s$, 
$\partial\mu_0/\partial\sigma=A/N\equiv a$, one obtains
\begin{equation}
\label{one-comp}
\Delta T=-\frac{a_I-a_{II}}{s_I-s_{II}}\Delta\sigma-\frac{c_I-c_{II}}{s_I-s_{II}}k_BT.
\end{equation}

Note that the coexistence temperature in the dilute solution is no longer 
determined by the surface tension alone, but by the amount of solute as 
well. That is, the coexistence line of the pure solvent in the $T,\sigma$ 
plane is, for the solution, drawn out into a sheet in the space of 
$T,\sigma$ and $\mu_s$, the solute chemical potential. Thus the change in 
transition temperature $\Delta T$ upon the addition of solute is only a 
meaningful quantity when the change, if any, of the independent 
thermodynamic variable, the surface tension, is specified.  For example, 
the miscibility transition temperature has been intentionally varied by 
controlling the surface tension explicitly \cite{portet12}. In the case in 
which the surface tension is held fixed, Eq. (\ref{one-comp}) reduces to
\begin{equation}
\label{shift1}
\Delta T=-\frac{c_I-c_{II}}{s_I-s_{II}}k_BT.
\end{equation}
The equation explains, {\em inter alia}, the observation 
\cite{ladbrooke68} that the addition of cholesterol to a one component 
membrane at constant tension causes a decrease in the transition 
temperature from liquid to gel phase. This follows because the cholesterol 
preferentially partitions into the liquid phase \cite{ladbrooke68} which 
has a larger entropy per particle than does the gel phase 
\cite{ladbrooke68}
 
The extension of the result of Eq. (\ref{one-comp}) to a membrane of $p$ 
components that acts as a solvent for the solute is straightforward. Let 
the membrane without solute have $N$ molecules of which $N_i=Nx_i$ are of 
component $ i=1...p.$ The total energy of any given phase can be written
\begin{eqnarray} U&=&TS+\sigma A+\sum_{i=1}^p\mu_iN_i,\nonumber\\
                             &=&TS+\sigma A+\sum_{i=1}^{p-1}(\mu_i-\mu_p)N_i+N\mu_p,\\
   {\rm with\ differential}\qquad                       dU&=&TdS+\sigma dA+\sum_{i=1}^{p-1}(\mu_i-\mu_p)dN_i+\mu_pdN,    
\end{eqnarray}
which leads to the Gibbs-Duhem equation
\begin{equation}
SdT+Ad\sigma+N\sum_{i=1}^{p-1}x_id(\mu_i-\mu_p)+Nd\mu_p=0.
\end{equation}

I again consider the thermodynamic potential
\begin{equation}
\Phi(T,\sigma,\{N_i\},N)=U-TS-\sigma A,\nonumber\\
\end{equation}
where $\{N_i\}$ denotes the set of $N_i, i=1,p-1$.
In the absence of solute,
\begin{eqnarray}
\Phi_0(T,\sigma,\{N_i\},N)&=&\sum_{i=1}^{p-1}N_i(\mu_{i,0}-\mu_{p,0})+N\mu_{p,0},\\
d\Phi_0&=&-SdT-Ad\sigma+\sum_{i=1}^{p-1}(\mu_i-\mu_p)dN_i+\mu_{p,0} dN,
\end{eqnarray}
and $\Phi_0$ is obtained from the partition function
\begin{equation}
\exp[-\Phi_0/k_BT]=Tr\prod_{i=1}^p\frac{1}{N_i!}\exp-[(H-\sigma A)/k_BT],
\end{equation}
where $H$ is the Hamiltonian of the multi-component system.

Again, let $n_s$ solute molecules be added to the system changing the 
thermodynamic potential to
$\Phi(T,\sigma,\{N_i\},N,n_s)$. Employing the same arguments as before for a dilute solution, one finds that the chemical potential
\begin{eqnarray}
\mu_p&=&\frac{\partial\Phi(T,\sigma,\{N_i\},N,n_s)}{\partial N},\nonumber\\
\label{change}
&=&\mu_{p,0}-k_BTc.
\end{eqnarray}
As there is nothing distinguishing the component $p,$ this is true for the 
chemical potentials of all components.

At coexistence of two phases, the chemical potentials of all components 
must be equal. It is convenient to consider $\mu_p$ a function of $T$, 
$\sigma$ and the $p-1$ independent chemical potential differences 
$\delta\mu_i\equiv \mu_i-\mu_p.$ Then the condition of coexistence can be 
written
\begin{equation}
\mu_{p,0}^I(T,\sigma,\{\delta\mu_i\})-k_BTc^I=\mu_{p,0}^{II}(T,\sigma,\{\delta\mu_i\})-k_BTc^{II}.
\end{equation}
Assume that in the absence of solute, the two phases are in coexistence so 
that
\begin{equation}
\mu_{p,0}^I(T_{0,co},\sigma_0,\{\delta\mu_{i,0}\})=\mu_{p,0}^{II}(T_{0,co},\sigma_0,\{\delta\mu_{i,0}\}).
\end{equation}
Now expand the temperature $T$ about $T_{0,co}$, the surface tension 
$\sigma$ about $\sigma_0$ and the chemical potential differences 
$\delta\mu_i$ about $\delta\mu_{i,0}$ to obtain the extension of Eq. 
(\ref{one-comp}),
\begin{eqnarray}
\label{final}
\Delta T&=&-\frac{1}{s^I-s^{II}}\left\{(a^I-a^{II})\Delta\sigma +\sum_{i=1}^{p-1} (x_i^I-x_i^{II})\Delta(\mu_i-\mu_p)+k_BT(c^I-c^{II})\right\}.
\end{eqnarray}
In the above $a^I-a^{II}$ is the difference in area per particle of the coexisting phases, $x_i^I-x_i^{II}$  the difference in mol fractions of component $i$ in the coexisting phases, and $c^I-c^{II}$ the difference in the mol fraction of the solute in the coexisting phases.
Note that the coexistence temperature is now a function of $p+1$ fields;  $\sigma,$ the $p-1$ chemical potential differences  $\{\mu_i-\mu_p\}$, and $\mu_s$ the solute chemical potential. These fields, or an equivalent number of conditions, must all be specified if the change in transition temperature upon the addition of solute is to be a meaningful quantity.

\begin{figure}[htbp]
\includegraphics[width=.5\columnwidth] {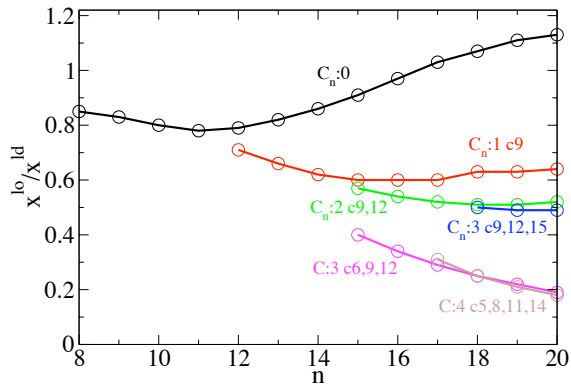}
\caption{Partition coefficient, $X^{lo}/X^{ld}$ for several kinds of single chains of length $n$. $C_n:0$ denotes a chain of length $n$ and no double bonds. From Ref. 4. }
\label{partition}
\end{figure}

The contribution to $\Delta T/T$ from the last term can be written
\begin{eqnarray}
-\frac{k_B}{s^I-s^{II}}\left(c^I-c^{II}\right)
&=&-\frac{k_B}{s^I-s^{II}}2{\bar c}\frac{1-c^{II}/c^I}{1+c^{II}/c^I},\nonumber\\
&=&-\frac{k_B}{s^I-s^{II}}2{\bar c}\frac{1-X^{II}/X^I}{1+X^{II}/X^I},\nonumber\\
&\approx&-\frac{k_B}{s^I-s^{II}}{\bar c }(1-X^{II}/X^I),
\end{eqnarray}
where ${\bar c}$ is the average solute concentration, and $X^I$ and 
$X^{II}$ are the mol fractions of the solute in the two phases. The last 
line follows when these mol fractions are not too different from one 
another.

Let phase $I$ be the liquid-disordered phase and $II$ be the 
liquid-ordered phase, in which case the entropy difference $s^I-s^{II}$ is 
positive, (see below). Then this contribution to $\Delta T/T$ is negative 
when the ratio $X^{II}/X^I$ is less than unity and is positive otherwise.  
The partition coefficient $X^{lo}/X^{ld}$ of several different kinds of 
chains in a bilayer consisting of dipalmitoyl phosphatidylcholine (DPPC), 
dioleoyl phosphatidylcholine (DOPC), and cholesterol were obtained 
recently\cite{uline10} from a self-consistent-field theory calculation, one that employed 
over $10^8$ configurations of each species of molecule in order to obtain partition 
functions. Fig. \ref{partition}, reproduced from that paper, shows results 
that are relevant here. The partition coefficients are plotted as a 
function of chain length, $n$. Note that for saturated chains, the 
partition coefficient decreases with increasing $n$ for small $n$, but for 
$n$ beyond 12 it increases with increasing $n$ and crosses unity for $n$ 
of about 16. The behavior is not difficult to understand. A saturated 
chain shorter than those which make up the bilayer partitions 
preferentially to the liquid disordered phase because its entropy is 
greater there\cite{degennes80}. This contribution dominates the energetic 
one which favors the liquid ordered phase. As $n$ increases to that of the 
saturated chains in the bilayer, the partition coefficient must take a 
value essentially equal to that of those chains. This follows from the 
fact that, if one added a lipid which was identical to one of the 
components of the bilayer, its partitioning into the two phases would simply be obtained from the 
endpoints of the tie line connecting them.  As saturated chains 
are found predominantly in the liquid ordered phase, the partition 
coefficient must exceed unity. As a consequence of this behavior of the 
partition coefficient, the contribution of the last term in Eq. 
(\ref{final}) would tend to cause the transition temperature to decrease 
upon the addition of octonal, and to decrease even more on the addition of 
decanol. But upon further increase of the chain length, the magnitude of 
the decrease in transition temperature would become smaller, and 
eventually vanish. This dependence of transition temperature on chain 
length is just the behavior observed by Gray et al. \cite{gray13}.
\section{Discussion}

I have shown that in small concentrations, the addition of a short-chain 
alcohol to a membrane undergoing a first-order transition to coexisting 
liquid-ordered, (lo), and liquid-disordered, (ld), phases causes a change 
in the transition temperature, as given by Eq. (\ref{final}); that of the 
several contributions to the change in transition temperature, one is 
proportional to the partitioning of the alcohol in the two phases; and 
that a recent calculation \cite{uline10} of this partitioning shows that 
this contribution would cause just the interesting behavior in the 
temperature shift as a function of chain length as is observed in 
experiment \cite{gray13}. Further, I now show that this term, and the 
observed order of magnitude of shift in the transition temperature, yields 
a reasonable difference in partitioning of the alcohol. To do so, I need 
the difference in entropy between ld and lo phases. This can be estimated 
from a combination of the Clausius-Clapeyron equation,
\begin{equation} 
\label{cc} 
\left.\frac{dT}{d\sigma}\right|_{coex}=-\frac{a^{I}-a^{II}}{s^{I}-s^{II}}, 
\end{equation} 
which gives the change in transition temperature with a change in surface 
tension, all other thermodynamic variables being fixed, the measured 
\cite{portet12} rate of change of transition temperature with surface 
tension, $dT/d\sigma\approx -2.8K/mN/m,$ and a difference in area per 
particle \cite{uline12,heftberger15} of $0.2nm^2.$ This yields a 
difference in entropy per particle of $(s^I-s^{II})/k_B\approx 5.2$. With 
this and a measured \cite{gray13} fractional decrease in transition 
temperature $\Delta T/T$ of about
$-0.013$, one obtains from Eq. (\ref{final}) a value 
$c^I-c^{II}\approx 0.1$ which is reasonable.

Note that the magnitude of the temperature shift given by Eq. 
(\ref{final}) depends upon the non-zero difference in entropy per particle 
in the two coexisting phases. Hence a calculation which assumes that this 
difference in entropy is zero, as is in a simple Ising model in which the 
entropy difference vanishes by symmetry \cite{meerschaert15}, cannot 
capture this temperature shift in a dilute solution.

I now address the question as to whether the contribution to the shift in 
transition temperature due to the partitioning of the solute, the last 
term in Eq. (\ref{final}), is the dominant one. The first term in Eq. 
(\ref{final}) can certainly be ignored compared to the last for the case 
of a biological membrane. The change in area per lipid 
\cite{uline12,heftberger15} between liquid-ordered and liquid-disordered 
phases is about $\Delta a=0.2 nm^2.$ Further, the surface tension 
decreases on the addition of solute, and this decrease cannot be larger 
than the surface tension itself.  In cells\cite{dai99}, this is on the 
order of $5\times 10^{-3}k_BT/nm^2.$ Thus in order for the first term in 
Eq. (\ref{final}) to be greater than the last, the difference in mol 
fractions of the solute in the two phases would have to be less than 
$1\times 10^{-3}.$

There remains to discuss only the terms in Eq. (\ref{final}) proportional 
to changes in chemical potential differences $\Delta(\mu_i-\mu_p),\ 
i=1,...p-1$. It would appear that these quantities are not controlled in 
the experiment, and to this extent, the change in transition temperature 
upon addition of alcohol is not a well-defined quantity; i.e. by varying 
these chemical potentials upon addition of the alcohol, one could vary the 
shift in transition temperature at will. Nevertheless it is reasonable to 
assume that, except for the addition of the short-chain alcohol, the 
composition of the giant plasma membrane vesicles utilized by Gray et al. 
\cite{gray13} are essentially the same as vesicles without alcohol. 
Therefore with the exception of the change in chemical potential of all 
solvent components brought about by the entropy of the solute, Eq. 
(\ref{change}), a change which does not affect the chemical potential 
differences $\mu_i-\mu_p$, the chemical potentials are otherwise 
unaffected. Thus the shifts, $\Delta(\mu_i-\mu_p)$, vanish. If this be the 
case, then Eq. (\ref{final}) reduces to
\begin{equation}
\label{approx}
\Delta T=-\frac{k_BT(c^I-c^{II})}{s^I-s^{II}}.
\end{equation}

The above calculation has determined the shift, on the addition of solute, to an onset temperature of transition from  a single phase to a region of two-phase coexistence,  as in the experiments of Gray et al.\cite{gray13}. I now briefly discuss the case in which there can be more than a single temperature of transition to consider. This situation is most simply discussed in the context of the liquid-gel transition in a one-component membrane. Were the 
surface tension to be held constant while the
temperature of the system in the liquid phase was reduced, then the system would enter the region of two-phase coexistence at a certain temperature; the transfer of liquid phase to gel phase would occur at the same temperature, and the system would emerge from the region of two-phase coexistence at this temperature.  The effect of adding a solute, such as cholesterol, on this transitionn temperature could then be calculated from Eq. (\ref{shift1}) and the shift would be unambiguous. However were the area, rather than the surface tension, to be fixed, then the system in the liquid phase would enter the two-phase region at a certain temperature, $T_1$, and the temperature would decrease while liquid phase was being converted to gel. Finally the system would emerge from the coexistence region and become pure gel at a temperature, $T_2$, lower than $T_1$. In this case the chemical potentials, $\mu_0(T_1,\sigma_1)$ and $\mu_0(T_2,\sigma_2)$ of the system would differ. Therefore upon the addition of cholesterol, the shifts $\Delta T_1$ and $\Delta T_2$ to the temperatures at which the coexistence region is entered and exited would differ.
However as the difference in the partitioning of cholesterol into the two phases would be expected to have the same sign at the two temperatures, and similarly for the difference in the specific entropies and areas, I would expect, from Eq. (\ref{one-comp}), that the shifts $\Delta T_1$ and $\Delta T_2$ would have the same sign even though their magnitudes would differ.  A similar argument can be made for a multicomponent membrane which exhibits a miscibility transition. If the external constraints were such that the temperature changes in the two-phase region as one phase is converted into the other, then one can expect that, upon the addition of a solute, the temperatures at which the two-phase region is entered and exited will be shifted by amounts of the same sign but of different magnitude. 
\section{Conclusions}

I conclude with a few observations and predictions. First I noted in the 
Introduction that the results of Gray et al.\cite{gray13} were 
interesting, {\em inter alia}, because the introduction of short-chain 
alcohols reduced the lo-ld miscibility transition whereas the addition of 
cholesterol caused it to increase. That is now readily understood from Eq. 
(\ref{approx}) and the fact that short-chain alcohols partition 
preferentially into the ld phase, that with the larger entropy per 
particle. Thus the signs of $c^I-c^{II}$ and $s^I-s^{II}$ are the same in 
Eq. (\ref{approx}). In contrast, cholesterol is known \cite{veatch03} to 
partition preferentially into the lo phase, the phase with the smaller 
entropy per particle. Thus $c^I-c^{II}$ and $s^I-s^{II}$ have opposite signs.

Second, the analysis presented here and the calculation of the partition 
coefficients shown in Fig. \ref{partition} predicts that if the addition 
of an alcohol with $n=14$ has almost no effect on the transition 
temperature, than the addition of an alcohol with $n=16$ will increase the 
transition temperature. This prediction has recently been confirmed 
\cite{machta16}.

Third, it can also be seen from Fig. \ref{partition} that the addition of 
alcohols with unsaturated bonds will lower the transition temperature more 
than those with saturated tails, and that for a given $n$ the magnitude of 
the decrease in transition temperature will increase with the degree of 
unsaturation.

I emphasize that the above analysis is relevant for first-order 
transitions of the solvent in which the solute concentrations, $c$, is 
sufficiently small that contributions quadratic in $c$ to the solvent 
chemical potential can be ignored.  How small this is can be estimated 
from the fact that the energy, being a homogeneous function of order 
unity, must depend upon the number of solute molecules, $n_s$, according 
to
\begin{equation}
U=\frac{1}{2}\frac{n_s^2}{n_s+N}J_1+\frac{n_sN}{n_s+N}J_2,
\end{equation}
where, as before, $N$ is the number of solvent molecules. The interaction 
strengths $J_1$ and $J_2$ are those between solute molecules themselves, 
and between solute and solvent molecules respectively. Differentiating 
with respect to $N$ we find the contribution to the chemical potential of 
solvent molecules is $c^2(J_2-J_1/2).$ Comparing this with the contribution 
to the solvent chemical potential which is of first order in the solute 
concentration, $-k_BTc$, Eq(\ref{mu-change}), we see that the arguments of 
this paper require that the solute concentration be less than $c^*\approx 
k_BT/J$ where $J$ is the order of magnitude of the larger of the two 
interaction strengths $J_1$ and $J_2$. If the concentration of solute is 
indeed less than $c^*$, then the analysis of this paper is applicable to 
first-order transitions, even those which are close to a critical point, 
as in the experiments of Gray et al.\cite{gray13}.

For concentrations larger than $c^*$ it is well-known that a solute which 
acts like an amphiphile, gaining energy by placing itself between the 
components of the solvent, decreases the miscibility transition 
temperature, while one that prefers {\it either} phase of the 
phase-separated system increases that temperature \cite{prigogine54}. 
These behaviors were manifest in a recent simulation \cite{meerschaert15}. 
Combining these results with those for the small concentrations of the 
dilute-solution regime, one sees that a solute which prefers the lo phase, 
the one with the smaller entropy per particle, will raise the transition 
temperature over a wide range of compositions. In contrast a solute which 
prefers the ld phase, that with the larger entropy per particle, will on 
first addition, decrease the transition temperature, but on further 
addition will eventually increase it. From this observation there results 
a fourth prediction: that a short-chain alcohol which, at small 
concentrations, had been observed to lower the miscibility transition 
temperature in a giant plasma membrane vesicle will actually raise that 
temperature if its concentration in the membrane can be increased 
sufficiently,

Finally I note that it has recently been observed \cite{cornell16} that 
short-chain alcohols added in small concentrations to three-component 
giant unilamellar vesicles raise the lo, ld miscibility transition 
temperature, in contrast to their behavior when added to the giant plasma 
membrane vesicles of Gray et al. \cite{gray13}. I would predict that, all 
other thermodynamic variables being held constant, smaller concentrations 
of short-chain alcohol than those used would lower the transition 
temperature in giant unilamellar vesicles. Of course I am assuming that 
the reduction in transition temperature resulting from this small 
concentration would be observable reliably. The difference between the 
results for the temperature shift in the two types of membranes could, 
perhaps, be related to the difference in their compositions which affects 
not only the partitioning of the solute into the coexisting lo and ld 
phases, but also the entropy per particle of those phases. Both of these 
factors, the latter particularly, affect the magnitude of the shift in 
transition temperature, as can be seen from Eq. (\ref{final}). Thus the 
temperature shift in giant unilamellar vesicles might be much smaller than 
in giant plasma vesicles. The difference in entropy per particle is, of 
course, directly related to the latent heat of the transition, so just how 
closely the behavior of the two different vesicles correspond to one 
another could be interrogated by calorometric methods.

I am grateful for the many rewarding discussions with Caitlin Cornell and 
Sarah Keller, and also thank Sarah Veatch and Mark Uline for useful 
correspondence.



\end{document}